\documentstyle[12pt,epsf]{article}
\newcommand{\nin}{\noindent}
\newcommand{\be}{\begin{equation}}
\newcommand{\ee}{\end{equation}}
\newcommand{\bea}{\begin{eqnarray}}
\newcommand{\eea}{\end{eqnarray}}

\newcommand{\nn}{\nonumber\\}

\newcommand{\ol}{\overline}

\begin{document}

\begin{center}

{\Large BUSSTEPP lecture notes

\vspace{0.5cm}

{\bf Exact Wilsonian Renormalization}}

\vspace{0.5cm}

J. Alexandre\\
Department of Physics, King's College London, WC2R 2LS, UK

\vspace{2cm}

{\bf Abstract}

\end{center}

\vspace{0.2cm}

These lecture notes introduce exact Wilsonian renormalisation, and describe its technical approach, 
from an intuitive implementation to more advanced realisations. The methods and concepts are 
explained with a scalar theory, and their extension to quantum gravity is discussed as an application.

\tableofcontents

\eject

\section{Motivations}

Wilsonian renormalisation provides an elegant way of building an effective theory, and gives an intuitive understanding of scale 
dependence in Quantum Field Theory (QFT). Its fundamental idea is based on the expectation that physics at large scale should 
be independent of most microscopic details, and predictions should involve a small portion only of all the parameters 
describing these details. As an example, the description of water flowing in a stream is independent of the details of the water molecule, and the corresponding
effective description is provided by Fluid Mechanics instead of Quantum Mechanics.

In QFT, because one deals with an infinite number of degrees of freedom, naive quantum corrections diverge, and one needs to regularise momentum integrals.
Any regularisation necessarily involves an energy scale which must be put by hand, and physical quantities then depend on 
this arbitrary scale.
The interpretation for this scale dependence is that a given system is described by different parameters at different energies. A would-be divergence is therefore  
turned into a scale dependence, which is the essence of the concept of renormalisation.

An intuitive understanding of this scale dependence originates from Statistical Mechanics, in the study of phase transitions, as explained below. 
The fundamental object, at the core of the method, is the partition function, whose properties allow for the derivation of exact functional identities. 
The corresponding Wilsonian approach to renormalisation in QFT is explained
in the present lecture notes, which focus on few essential points, and more detailed reviews can be found in \cite{lectures}.

\subsection{Scale dependence of a theory}

In the 4-dimensional scalar theory with interaction $\phi^4$, the one-loop coupling $g^{(1)}$ is formally given by
\be
ig^{(1)}=ig_b+\frac{3\hbar(ig_b)^2}{2}\int\frac{d^4p}{(2\pi)^4}\frac{i^2}{(p^2-m^2)^2}~,
\ee
where $g_b$ is the bare coupling, the factor 3 arises from the three possibilities of displaying the external lines, and 1/2 is 
the symmetry factor of the graph. A Wick rotation leads to
\be
g^{(1)}=g_b-\frac{3\hbar g_b^2}{32\pi^2}\int \frac{xdx}{(x+m^2)^2}~,
\ee
which is logarithmically divergent. To avoid this divergence, one can introduce by hand the ultraviolet (UV) cut off $\Lambda$, which leads to
\be\label{g1loop}
g^{(1)}=g_b-\frac{3\hbar g_b^2}{16\pi^2}\ln\left(\frac{\Lambda}{m}\right)~+~\mbox{finite}~=g_b-\frac{3\hbar (g^{(1)})^2}{16\pi^2}\ln\left(\frac{\Lambda}{m}\right)+{\cal O}(\hbar^2)~.
\ee
The process of regularization of the loop integral therefore introduces a mass scale, such that a potential divergence has been replaced 
by a scale dependence. If one uses dimensional regularization instead, the arbitrary scale is introduced in the bare coupling 
$g_b\Lambda^\epsilon$, which has a positive mass dimension in space time dimension $d=4-\epsilon$. But
in any case, one needs to introduce an arbitrary mass scale $\Lambda$ if one wishes to avoid the divergence.\\
One can obtain a $\Lambda$-independent quantity though: 
the beta function $\beta\equiv\Lambda\partial_\Lambda g_b$ at fixed $g^{(1)}$, which at one loop is 
\be
\beta^{(1)}=\frac{3\hbar (g^{(1)})^2}{16\pi^2}~.
\ee
The scale-dependence obtained from the regularization of loop integrals might seem artificial, but is actually a deep feature of
QFT, which can be understood in a more intuitive way through the elegant process of Wilsonian renormalisation.

\subsection{Spin blocks}

Wilson's idea of renormalisation was originally motivated by the study of condensed matter systems in the vicinity of
a phase transition. Let's take the example of a ferromagnetic sample, where two effects are competing: \\
{\it(i)} Magnetic order, as a result of the interactions between spins located on a lattice, tending to align all the 
spins in the same direction. The situation where this effect dominates corresponds to the spontaneous symmetry breaking phase, 
where the specific direction of spins breaks the rotation group $O(3)$ to $U(1)$. The total magnetisation, which plays the 
role of order parameter, is non-vanishing;\\ 
{\it(ii)} Thermal disorder, which tends to give spins random directions. If this effect dominates, the system is in the 
symmetric phase, where the total magnetisation vanishes.

The situation where these two effects are of the same order corresponds to the phase transition, and the correlation between spins 
becomes important: the system is very sensitive to the modification in the direction of one spin, which is felt by other spins 
located many lattice sites away. The phase transition therefore involves many degrees of freedom, interacting with each other.

The concept of spin blocks is motivated by the idea that, when the correlation length $\xi$ becomes large 
(compared to the lattice spacing), details with a typical size $<<\xi$ should not play a role
in the features of the phase transition, such that these details can be integrated out in order to simplify the description of the system.
Integrating out spins $S_i^{(0)}$ can be achieved by defining new spin variables $S^{(1)}_j$ from a block of the original spins.
The new Hamiltonian $H^{(1)}[S^{(1)}]$ of the system can then be expressed in terms of the new spin variables as follows
\be
\exp(-H^{(1)}[S^{(1)}])=\sum_{S^{(0)}}\delta\Big(S^{(1)}-f(S^{(0)})\Big)\exp(-H^{(0)}[S^{(0)}])~,
\ee
where $H^{(0)}$ is the original Hamiltonian, defined with the original spin variables $S^{(0)}$, and $f$ corresponds to the definition
of the blocks, which contains some original spins. The partition function of the system is independent of the spin variables
\be
Z=\sum_{S^{(0)}}\exp(-H^{(0)}[S^{(0)}])=\sum_{S^{(1)}}\exp(-H^{(1)}[S^{(1)}])~,
\ee
which leads to the same physical predictions. The blocking procedure can be repeated again and again, leading to a chain of Hamiltonians
$\{H^{(n)}\}$, each defined by a set of parameters which depend on the blocking step $n$. This construction therefore provides a 
scale-dependent description of the system. The simplification in the description occurs because, among the potentially large
set of parameters defining the Hamiltonian, many will play no role in the infrared (IR) limit, where the system is zoomed out 
(irrelevant parameters), and only few of them will dominate the IR description (relevant parameters). The scale dependence of parameters
generates renormalisation flows, which are discussed below in the context of scalar field theory.

\subsection{One-loop Wilsonian renormalisation flows}

We work in Euclidean space; the IR field is denoted $\phi$ and has non-vanishing Fourier components for $p\le k$;
the UV field is denoted $\psi$ and has non-vanishing Fourier components for $k<p\le\Lambda$.\\
The original action, defined at some scale $\Lambda$, is of the form
\be
S_\Lambda[\Phi]=\int_x\left(\frac{1}{2}\partial_\mu\Phi\partial^\mu\Phi+U_\Lambda(\Phi)\right)~,
\ee
where $\Phi=\phi+\psi$, and the effective action $S_k$ at the scale $k$ is defined by 
\be\label{blocking}
\exp\left(-\frac{1}{\hbar}S_k[\phi]\right)=\int{\cal D}[\psi]\exp\left(-\frac{1}{\hbar}S_\Lambda[\phi+\psi]\right)~.
\ee
This definition of effective action corresponds to defining ``block spins'' with lattice spacing $k^{-1}$ from the original lattice spacing $\Lambda^{-1}$.   
If we are interested in the one-loop effective potential only, it is enough to consider a uniform IR field $\phi=\phi_0$, such that 
the action is $S_k[\phi_0]=VU_k(\phi_0)$, where $U_k$ is the running potential and
$V$ is the space time volume, also equal to 
\be
V\equiv\int d^4x=\left.\int d^4x\exp(ip x)\right|_{p=0}=\tilde\delta(0)~.
\ee
The next step is to expand the action $S_\Lambda[\phi_0+\psi]$ about $\phi_0$. The first functional derivative of $S$ does not feature in this expansion, since 
the fluctuation field has Fourier modes for non-vanishing momentum only, and $(\delta S/\delta\phi)_{\phi_0}\propto\tilde\delta(0)$, such that
\be
\int_p\tilde\psi_p\left.\frac{\delta S}{\delta\phi}\right|_{\phi_0}=0~.
\ee
The blocking (\ref{blocking}) leads to
\bea
&&\exp\left(-\frac{V}{\hbar}U_k(\phi_0)\right)\\
&=&\exp\left(-\frac{V}{\hbar}U_\Lambda(\phi_0)\right)
\int{\cal D}[\psi]
\exp\left(-\frac{1}{2\hbar}\int_{k\le|p|\le\Lambda}[p^2+U_\Lambda''(\phi_0)]\tilde\psi_p\tilde\psi_{-p}+\cdots\right)~,\nonumber
\eea
where dots represent higher order in $\psi$, which contribute to higher orders in $\hbar$ (this can be seen with the change of functional variable
$\psi\to\sqrt\hbar~\psi$).
The resulting Gaussian integral is calculated using
\be
\int{\cal D}[\psi]\exp\left(-\tilde\psi_p{\cal O}_{pq}\tilde\psi_q\right)=\frac{1}{\sqrt{\mbox{det}{\cal O}_{pq}}}
=\exp\left(-\frac{1}{2}\mbox{Tr}\left\{\ln{\cal O}_{pq}\right\}\right)~,
\ee
as well as the logarithm and the trace of a diagonal operator
\bea
\ln[F(p)\tilde\delta(p+q)]&=&\tilde\delta(p+q)\ln[F(p)]\\
\mbox{Tr}\left\{G(p)\tilde\delta(p+q)\right\}&=&\int_p\int_q\tilde\delta(p+q)G(p)\tilde\delta(p+q)=V\int_p G(p)~.
\eea
In the present case, Fourier modes integrated out are defined for $k<|p|\leq\Lambda$, such that
\bea\label{blockingbis}
U_k(\phi_0)&=&U_\Lambda(\phi_0)+\frac{\hbar}{2V}\mbox{Tr}_{k<|p|\le\Lambda}\left\{\tilde\delta(p+q)\ln[p^2+U_\Lambda''(\phi_0)]\right\}
+{\cal O}(\hbar^2)\nn
&=&U_\Lambda(\phi_0)+\frac{\hbar}{2}\int_k^\Lambda \frac{d^4p}{(2\pi)^4}
\ln\left(\frac{p^2+U_\Lambda''(\phi_0)}{p^2+U_\Lambda''(0)}\right)+{\cal O}(\hbar^2)~,
\eea
where the origin of the potential is chosen such that $U_k(0)=0$. 
A derivative of eq.(\ref{blockingbis}) with respect to $k$ finally leads to the one-loop flow equation
\be\label{1loopflow}
k\partial_k U_k(\phi_0)=-\frac{\hbar k^4}{16\pi^2}\ln\left(\frac{k^2+U_\Lambda''(\phi_0)}{k^2+U_\Lambda''(0)}\right)+{\cal O}(\hbar^2)~,
\ee
which shows the explicit scale dependence of the effective potential defined at the scale $k$.

\section{Exact flows (sharp cut off)}

The one-loop flow equation (\ref{1loopflow}) is enough when quantum effects are perturbative only, and higher-orders in $\hbar$ can
indeed be neglected. If one considers non-perturbative effects though (as in section 3.3 for example), 
it is necessary to consider an improved Wilsonian renormalisation procedure. This can be obtained by lowering the cut off infinitesimally,
as explained here.

\subsection{Wegner-Houghton equation}

We consider here the local potential approximation where, for all $k$, the running action has the form
\be\label{locpot}
S_k[\phi]=\int_x\left(\frac{1}{2}\partial_\mu\phi\partial^\mu\phi+U_k(\phi)\right)~.
\ee
This corresponds to a projection of the action on a subspace of functional space, where only the non-derivative part of the action 
is allowed to evolve with $k$, but not the kinetic term, higher order derivatives or derivative interactions.\\

\nin Instead of integrating Fourier modes from the cut off $\Lambda$ to some scale $k$ in one go, 
we start the blocking procedure from $k$, and implement an infinitesimal step $dk<<k$. We now eliminate Fourier modes $\psi_p$
which are non-zero for $k-dk<|p|\le k$:
\bea
&&\exp\left(-\frac{V}{\hbar}U_{k-dk}(\phi_0)\right)\\
&=&\exp\left(-\frac{V}{\hbar}U_k(\phi_0)\right)
\int{\cal D}[\psi]\exp\left(-\frac{1}{2\hbar}\int_{k-dk<|p|\le k}[p^2+U_k''(\phi_0)]\tilde\psi_p\tilde\psi_{-p}+\cdots\right)\nonumber
\eea
where, this time, higher orders in $\psi$ involve higher orders in $dk$, since the trace is 
taken in the infinitesimal shell of radius $k$ and thickness $dk$. As a consequence, we get
\bea
U_{k-dk}(\phi_0)&=&U_k(\phi_0)+\frac{\hbar}{2V}\mbox{Tr}_{k-dk\le|p|\le k}\left\{\tilde\delta(p+q)\ln[p^2+U_k''(\phi_0)]\right\}
+{\cal O}(dk/k)^2\nn
&=&U_k(\phi_0)+\frac{\hbar 2\pi^2 k^3 dk}{2(2\pi)^4}\ln\left(\frac{k^2+U_k''(\phi_0)}{k^2+U_k''(0)}\right)+{\cal O}(dk/k)^2~,
\eea
where $\tilde\delta(0)=V$ is used.
The limit $dk\to0$ leads then to the equation satisfied by the running potential
\be\label{WH}
k\partial_k U_k(\phi_0)=-\frac{\hbar k^4}{16\pi^2}\ln\left(\frac{k^2+U_k''(\phi_0)}{k^2+U_k''(0)}\right)~.
\ee
This exact equation was initially derived in \cite{WegnerHoughton}, and is self-consistent: the running potential appears on 
both sides, the Wegner-Houghton equation is a bit similar to a differential Schwinger-Dyson equation. For this reason, it
consists in a partial resummation of all the orders
in $\hbar$. The resummation is partial because eq.(\ref{WH}) is derived in the framework of the approximation (\ref{locpot}).
If one expands the potential in powers of $\hbar$, then $U_k(\phi_0)=U_\Lambda(\phi_0)+{\cal O}(\hbar)$, and the Wegner Houghton equation
(\ref{WH}) gives the one-loop flow equation (\ref{1loopflow}). \\

\subsection{Fixed point and classification of coupling constants}

\nin One then parametrise the running potential in the polynomial form
\be
U_k(\phi_0)=\frac{g_2(k)}{2}\phi_0^2+\frac{g_4(k)}{24}\phi_0^4+\frac{g_6(k)}{6!}\phi_0^6+\cdots
\ee
An expansion of the right-hand side of the flow equation (\ref{WH}) in powers of $\phi_0$ generates an infinite series of terms, 
that we truncate here to $\phi_0^6$. 
The identification of powers of $\phi_0$ of both sides of the equation leads then to
\bea
k\partial_kg_2(k)&=&-\frac{\hbar k^4}{16\pi^2}~\frac{g_4(k)}{k^2+g_2(k)}\\
k\partial_kg_4(k)&=&-\frac{\hbar k^4}{16\pi^2}
\left(\frac{-3g_4^2(k)}{[k^2+g_2(k)]^2}+\frac{g_6(k)}{k^2+g_2(k)}\right)\\
k\partial_kg_6(k)&=&-\frac{15\hbar k^4}{16\pi^2}
\left(\frac{2g_4^3(k)}{[k^2+g_2(k)]^3}-\frac{g_4(k)g_6(k)}{[k^2+g_2(k)]^2}\right)~.
\eea
The dimensionless couplings $\tilde g_{2n}$ are, in 4 dimensions,
\be
g_{2n}(k)=k^{4-2n}\tilde g_{2n}~,~~~~~\mbox{with}~~[\tilde g_{2n}]=0~, 
\ee
which lead to the renormalisation equations 
\bea\label{flow1}
k\partial_k \tilde g_{2}&=&-2\tilde g_2-\frac{\hbar}{16\pi^2}~\frac{\tilde g_4}{1+\tilde g_2}\\
k\partial_k\tilde g_4&=&~~0~~~-\frac{\hbar}{16\pi^2}\left(\frac{-3\tilde g_4^2}{[1+\tilde g_2]^2}+\frac{\tilde g_6}{1+\tilde g_2}\right)\\
k\partial_k \tilde g_6&=&~~2\tilde g_6~-\frac{15\hbar}{16\pi^2}\left(\frac{2\tilde g_4^3}{[1+\tilde g_2]^3}
-\frac{\tilde g_4\tilde g_6}{[1+\tilde g_2]^2}\right)~,
\eea
where $\tilde g_4=g_4$, since $[g_4]=0$.\\
By definition, a fixed point $\tilde g^\star=\{\tilde g^\star_{2n}\}$ of the renormalisation flows (\ref{flow1}) 
 is invariant in the blocking procedure: $\partial_k \tilde g^\star=0$. Once a fixed point is found,
one can classify the coupling constants $g\ne g^\star$ according to their behaviour when $k\to0$:\\
{\it Relevant coupling}: $\tilde g_{2p}$ goes away from $\tilde g_{2p}^\star$ as $k\to0$\\
{\it Irrelevant coupling}: $\tilde g_{2q}$ converges to $\tilde g_{2q}^\star$ as $k\to0$\\

\nin In the previous example, the only fixed point is the trivial one: $\tilde g^\star=0$. 
This is called the Gaussian fixed point because the action contains the kinetic term only, which is quadratic in the field.
Compared to this trivial fixed point, one can classify the coupling constants (the classical scaling - if not zero - is dominant):\\
$\tilde g_2(k)$ is relevant, since it increases in the IR: $\partial_k \tilde g_{2}<0$;\\
$\tilde g_6(k)$ is irrelevant, it decreases in the IR: $\partial_k \tilde g_6>0$.\\
If one truncates the theory to $\phi_0^4$, then $\partial_k \tilde g_4>0$: $\tilde g_4$
is irrelevant. Note that this is a consequence of quantum fluctuations only, since at the classical level ($\hbar=0$), $\tilde g_4$
is {\it marginal} = it does not depend on $k$.

\vspace{0.5cm}

\nin Once coupling constants are classified, one defines a {\it universality class} as a set of theories which differ only by irrelevant 
parameters. In this case, the renormalisation flows of these different theories all lead to the same IR physics, defined by the set
of relevant parameters, because a modification to an irrelevant parameter will not have any consequence in the IR.

\vspace{1cm}

\nin {\bf Exercise 1}: {\it Critical exponents}\\
A critical exponent $\alpha$ is defined by the power law $\xi=\xi_0(k/k_0)^\alpha$ which gives the evolution of the parameter $\xi$ with the scale $k$.
Consider a model defined by the dimensionless couplings $\tilde g=\{\tilde g_1,\tilde g_2,\cdots\}$ and the renormalisation group equations 
$k\partial_k\tilde g=f(\tilde g)$, with the fixed point $\tilde g^\star$.
After linearising the renormalisation equations, 
show that the eigenvalues of the matrix $M_{ab}\equiv\partial f_a/\partial g_b|_{\tilde g^\star}$ are the critical exponents of the model.\\

\nin {\bf Exercise 2}: {\it Wilson-Fisher fixed point}\\
Write the Wegner-Houghton equation in dimension $d=4-\epsilon$ and the corresponding evolution equations of the dimensionless couplings, after truncating
the equation to $\phi^4$. Show that there is a non-trivial fixed point, obtained by expanding the flow equations to the quadratic order in the running couplings.

\subsection{Relation to renormalisability}

The blocking procedure defines renormalisation flows which go from high momenta to the IR. On the other hand, in Particle Physics, 
one is interested in the high-energy behaviour, with a fixed scale $\Lambda$ and $k\to\infty$,
such that a relevant coupling corresponds to a super-remormalisable parameter: it decreases in the UV and its behaviour is controlled.
On the other hand, an irrelevant coupling corresponds to a non-renormalisable parameter: 
it increases in the UV and diverges as $k\to\infty$.\\

\nin A classically marginal coupling corresponds to a renormalisable theory, and the behaviour of the coupling at high energy
depends on the sign of quantum fluctuations to the renormalisation flow. In the situation of the renormalisable $\phi^4$ bare theory,
quantum fluctuations make $g_4$ increase with $k$:
if one truncates the theory to $\phi^4$, one obtains, for high momentum, the equation
\be
k\partial_k g_4(k)=\frac{3\hbar}{16\pi^2}~\frac{g_4^2(k)}{[1+g_2(k)/k^2]^2}\simeq\frac{3\hbar}{16\pi^2}~g_4^2(k)~,
\ee
with solution
\be
g_4(k)=g_4(\Lambda)\left(1-\frac{3\hbar}{16\pi^2}g_4(\Lambda)\ln\left(\frac{k}{\Lambda}\right)\right)^{-1}~.
\ee
One can make two comments at this point:
\begin{itemize} 
\item The latter solution corresponds to the resummation of a geometric series of $n$-loop graphs, corresponding to the product of $n$
one-loop graphs (\ref{g1loop}). The present truncation of the renormalisation equation (\ref{WH}) 
therefore provides us with an improved one-loop calculation;
\item In the spirit of Wilsonian renormalisation, $k\le\Lambda$, such that $g_4(k)$ is never singular. 
But if one fixes the scale $\Lambda$ and
increases $k$, a singularity occurs at
\be
k_\infty=\Lambda\exp\left(\frac{16\pi^2}{3\hbar g_4(\Lambda)}\right)>>\Lambda~.
\ee
This singularity also occurs in QED, where it corresponds to the {\it Landau pole} 
\be
k_{Landau}\simeq m_{electron}\exp(685)>>M_{Planck}~.
\ee
In QCD though, the coupling decreases with $k$, such that the 
theory is asymptotically free.
\end{itemize}

\subsection{Maxwell construction}

In the situation of a bare potential containing a concave part (as for the double-well potential), non-trivial saddle points
in $\psi$ appear at some stage of the blocking procedure \cite{ABP}, in order to avoid the ``spinodal instability'' which would 
otherwise occur at the scale $k_s$ satisfying $k_s^2+U_{k_s}''(\phi_0)=0$, when the restoration force for quadratic 
fluctuations vanish. These saddle points 
lead to the Maxwell construction in the limit $k\to0$, where the effective potential if flat between the bare minima. 
This flattening ensures that the effective potential is convex, as expected from 
general arguments (see Appendix B). Indeed, in the limit of infinite volume, the Wilsonian effective potential is identical 
to the one-particle irreducible (1PI) effective potential (see Appendix C), and must therefore be convex in the IR limit $k=0$.

\begin{figure}
\epsfxsize=8cm
\centerline{\epsffile{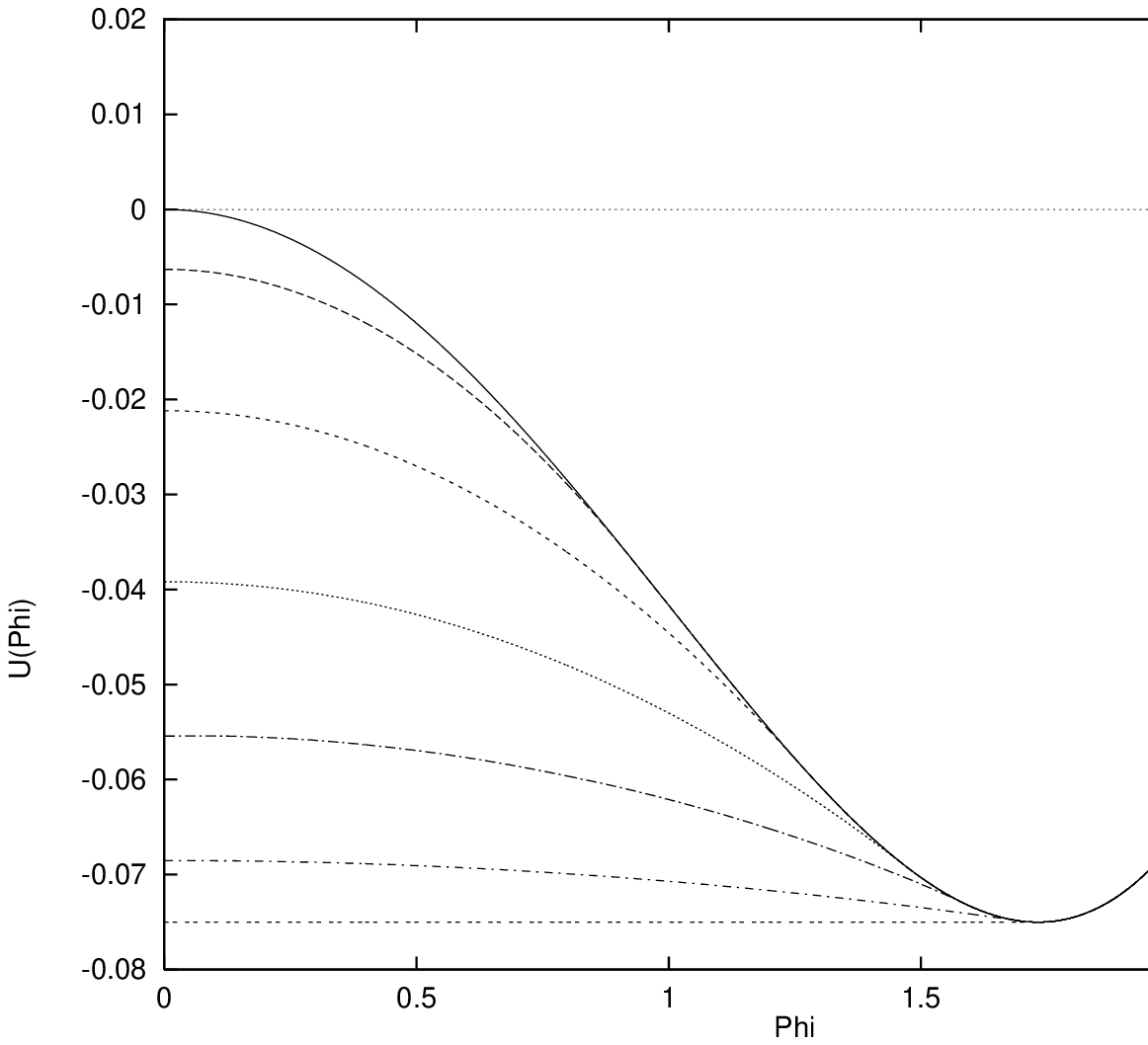}}
\caption{Evolution of the running potential in the situation of a concave bare potential. Convexity is recovered in the 
IR limit $k\to0$, where the effective potential becomes flat, as expected from the Maxwell construction. (Figure taken from \cite{ABP}) }
\end{figure}

Taking into account the presence of a non-trivial saddle point $\psi_{s}$ in the integration over the shell of thickness $dk$, 
the blocking reads
\bea
&&\exp\left(-\frac{V}{\hbar}U_{k-dk}(\phi_0)\right)=\exp\left(-\frac{1}{\hbar}S_k[\phi_0+\psi_s]\right)\\
&\times&\int{\cal D}[\psi]
\exp\left(-\frac{1}{2\hbar}\int_p\frac{\delta^2S_\Lambda[\phi_0+\psi_s]}{\delta\phi_p\delta\phi_{-p}}(\psi-\psi_s)_p(\psi-\psi_s)_{-p}
+\cdots\right)~.\nonumber
\eea
Ignoring quantum fluctuation, the latter blocking gives
\be
VU_{k-dk}(\phi_0)\simeq S_k[\phi_0+\psi_s]~,
\ee
which leads to a finite-difference equation for the running potential $U_k$. This ``tree-level'' renormalisation flow has been studied 
numerically in \cite{ABP}: the saddle point is assumed to be a plane wave, whose amplitude is evaluated at each blocking step, and contributes to the gradual elimination of 
the concave part of the bare potential (see Fig.1). 
Note that this convexity is not obtained if the scalar field is coupled to a gauge field, as in the Standard Model: in order
to define the partition function, one needs then to fix a gauge and therefore restrict the field space over which the path integral 
is defined. This restriction imposes to quantise the theory over one given vacuum, whereas convexity is obtained by taking into account
the different vacua of the bare theory, when defining the partition function. A analytical derivation of the convex 1PI effective potential is given in 
\cite{AT}, where the Maxwell construction is obtained in the limit of infinite volume: the flat dressed potential describes a system where the ground state is a 
superposition of the two bare vacua.

\subsection{Problems with a sharp cut off}

The description given in this section has the advantage of being intuitive and free of any additional technicality.
The sharp cut off used here has two important problems though: it cannot be used for a theory including gauge invariance, and
it cannot predict the evolution of the derivative terms of the action.

As far as gauge invariance is concerned, one cannot imposes a cut off $\Lambda$ on Fourier components of a gauge field:
a gauge transformation would spoil this cut off, since it involves any gauge function, which 
can have non-vanishing Fourier modes for momentum larger than $\Lambda$.

The point concerning the derivative terms in the action is more subtle. In the previous example, a constant IR field has been used, which is enough to 
evaluate the potential part of the action (the part containing no derivatives). In order to derive the evolution of derivative terms, one needs to 
consider a coordinate-dependent IR field, for example 
\be\label{phi1}
\phi=\phi_0+\phi_1\sin(p^\mu x_\mu)~,
\ee
where $p$ is fixed (with $|p|<k$) and $\phi_1$ is a constant. 
Each blocking step introduces an infinite series of derivative terms, and the running action can be parametrised by
the derivative expansion
\be\label{gradexp}
S_k[\phi]=\int d^4x\left\{\frac{1}{2}\sum_{n=0}^\infty Z_k^{(n)}(\phi)\partial_\mu\phi\Box^n\partial^\mu\phi+U_k(\phi)\right\}~,
\ee
where $Z_k^{(n)}(\phi)$ are functions allowing wave function renormalisation and derivative interactions which are introduced by the blocking.
The different evolution equations are then obtained, from the renormalisation equation, by the identification of the
following terms:\\
$\bullet$ terms independent of $\phi_1$ for the potential $U_k(\phi)$;\\
$\bullet$ terms proportional to $p^{2n+2}\phi_1^2$ for $Z^n_k(\phi)$.\\
(Terms depending on $\phi_1$ but not on $p$ give the evolution equations for the field derivatives $\partial_\phi^n U_k(\phi)$, which are consistent with
the evolution for $U_k(\phi)$.)\\
For the IR field (\ref{phi1}), the second derivative of the action (\ref{gradexp}) is of the form
\be
\frac{\delta^2S_k}{\delta\phi_x\delta\phi_y}=F(\sin(p x),\partial_\mu)\delta(x-y)~,
\ee
such that the evolution equation contains terms of the form
\be
\int_{\cal D}\tilde F(p,q)\tilde\psi_q\tilde\psi_{-p-q}~.
\ee
Because one integrates Fourier modes $\tilde\psi$ in the shell of radius $k$ and thickness $dk$, the domain ${\cal D}$ of integration over $q$
of the latter integral is deformed and doesn't correspond to the spherical shell any more: ${\cal D}$ is defined by the simultaneous conditions
\be
k-dk\le|q|\le k~,~~\mbox{and}~~k-dk\le|q+p|\le k~,
\ee
which can be achieved for all $q$ only if $|p|<<dk$. But in order to obtain the flow equation, one takes the limit $dk\to0$,
which leads then to  $\phi=\phi_0$, and therefore we are left with the original situation where only the evolution 
of the potential can be found.

\section{Exact flows (smooth cut off)}

We explain here how both problems with the Wegner-Houghton approach can be avoided, by introducing a cut off function which 
allows a progressive (``smooth'') elimination of Fourier modes.
In this section and the following, we set $\hbar=1$, and Fourier modes are denoted without a tilde.

\subsection{Polchinski renormalisation equation}

The idea of a smooth cut off was introduced by Polchinski \cite{Polchinski}. 
High momentum Fourier modes are gradually cut off above a given scale $k$, by replacing the inverse propagator $p^2+m^2$ with 
a differentiable cut-off function $Q_k^{-1}(p^2)$ which satisfies the following conditions
\begin{itemize}
\item $Q_k^{-1}(p^2)= p^2+m^2$ for $p^2\le k^2$, such that Fourier modes for $p^2\le k^2$ propagate as expected;
\item $Q_k(p^2)$ decreases rapidly to 0 for $p^2>k^2$, such that Fourier modes for $p^2>k^2$ dominate the path integral and are thus preferentially integrated out.
\end{itemize}
We are interested in the evolution of the running action $S_k$, describing Fourier modes for $|p|\leq k$.  
The total running action, including the cut off function and the source term, is
\be
\Sigma_k=S_k+\frac{1}{2}\int_p\phi_{-p}Q_k^{-1}\phi_p+\int_pj_{-p}\phi_p~,
\ee
and the scale dependence of the action $S_k$ is obtained by imposing that the partition function is independent of $k$.
\nin The source $j_p$ is assumed to vanish for $p^2>k^2$, such that we have 
\be\label{jQ}
j_p\partial_kQ_k^{-1}(p^2)=0~,
\ee
and the partition function is
\be
Z[j]=\int{\cal D}[\phi]\exp\left(-\Sigma_k[\phi,j]\right)~.
\ee
The IR physics should be independent of the arbitrary scale $k$, which implies
\be\label{partialkZ}
\partial_kZ=0=-\int{\cal D}[\phi]\left(\partial_kS_k[\phi]+\frac{1}{2}\int_p\phi_{-p}\partial_kQ_k^{-1}\phi_p\right)\exp\left(-\Sigma_k[\phi,j]\right)~.
\ee
The next step is to find what the variation $\partial_kS_k$ should be, to satisfy $\partial_kZ=0$. For this, 
we note that the following total functional derivative can be written
\bea\label{exo2}
&&\partial_kQ_k^{-1}\left[\frac{\delta^2e^{-\Sigma_k}}{\delta\phi_p\delta\phi_{-p}}
+Q_k^{-1}\frac{\delta(\phi_pe^{-\Sigma_k})}{\delta\phi_p}
+Q_k^{-1}\frac{\delta(\phi_{-p}e^{-\Sigma_k})}{\delta\phi_{-p}}\right]\\
&=&Q_k^{-2}\left[Q_k\partial_k Q_k^{-1}\tilde\delta(0)-\phi_{-p}\partial_kQ_k^{-1}\phi_p-\partial_kQ_k
\left(\frac{\delta S_k}{\delta\phi_p}\frac{\delta S_k}{\delta\phi_{-p}}-\frac{\delta^2S_k}{\delta\phi_p\delta\phi_{-p}}\right)
\right]e^{-\Sigma_k}~,\nonumber
\eea
where the condition (\ref{jQ}) was used. We therefore see that, after ignoring the field-independent term 
$Q_k\partial_k Q_k^{-1}\delta(0)$, if we choose
\be\label{polchequa}
\partial_k S_k\equiv\frac{1}{2}\int_p\partial_kQ_k(p^2)\left(\frac{\delta S_k}{\delta\phi_p}\frac{\delta S_k}{\delta\phi_{-p}}
-\frac{\delta^2S_k}{\delta\phi_p\delta\phi_{-p}}\right)~,
\ee
the scale-independence condition (\ref{partialkZ}) is satisfied: the functional integral of the functional derivative vanishes, since the integrand 
decreases exponentially for large field amplitudes.
The self-consistent equation (\ref{polchequa}) describes how the bare action $S_k$ should evolve, 
for the IR physics to be unchanged when $k$ varies.\\

\nin Exercise 4 consists in solving the Plochinski equation in a simple context, where only quadratic and quartic terms in the field are taken into account, but where
any power of the momentum is allowed. On the other hand, the local potential approximation consists in allowing any function of the field for the running potential, 
but neglecting any correction to momentum-dependent parts of the action. For the Polchinski equation, this consists in projecting the running action on the functional subspace 
\be
S_k=\int_x~U_k(\phi)~~~~~\mbox{for all}~k~,
\ee
such that, for a constant IR configuration $\phi_0$,
\bea
\frac{\delta S_k}{\delta\phi_p}&=&\int_x\frac{\delta S_k}{\delta\phi_x}\frac{\delta\phi_x}{\delta\phi_p}=\int_xU_k'(\phi_0)e^{ipx}=U_k'(\phi_0)\tilde\delta(p)\\
\frac{\delta^2 S_k}{\delta\phi_p\delta\phi_q}&=&\int_x\int_y\frac{\delta^2 S_k}{\delta\phi_x\delta\phi_x}\frac{\delta\phi_x}{\delta\phi_p}\frac{\delta\phi_y}{\delta\phi_q}
=\int_x\int_yU_k''(\phi_0)\delta(x-y)e^{ipx+iqy}=U_k''(\phi_0)\tilde\delta(p+q)~.\nonumber
\eea
For this constant IR configuration $\phi_0$, we also have $S_k=VU_k(\phi_0)$, such that the Polchinski equation gives
\bea
\partial_kU_k(\phi_0)&=&\frac{1}{2V}[U_k'(\phi_0)]^2\int_p\partial_kQ_k(p^2)[\tilde\delta(p)]^2-\frac{1}{2V}U_k''(\phi_0)\int_p\partial_kQ_k(p^2)\tilde\delta(0)\nn
&=&\frac{1}{2}[U_k'(\phi_0)]^2\partial_kQ_k(0)-\frac{1}{2}U_k''(\phi_0)\int_p\partial_kQ_k(p^2)~.
\eea
Although there is no logarithm as in the Wegner-Houghton equation, it is here the quadratic term $[U_k']^2$ which leads to a non-trivial renormalisation flow.

\vspace{1cm}

\nin{\bf Exercise 3}: Derive eq.(\ref{exo2})\\

\nin{\bf Exercise 4}: {\it Quartic ansatz for the Polchinski equation}\\
Assume the following quartic ansatz  
$$
S_k=\frac{1}{2}\int_p\phi_pF_k(p^2)\phi_{-p}+\frac{1}{24}\int_{pqr}G_k(p,q,r)\phi_p\phi_q\phi_r\phi_{-p-q-r}~,
$$
and derive from eq.(\ref{polchequa}) the evolution equations for the functions $F$ and $G$.

\subsection{Wetterich average effective action}

An elegant way to implement exact Wilsonian renormalisation is through the average 
effective action, introduced by Wetterich (see \cite{Wetterich} for a review),
which corresponds to a 1PI effective action for which modes with $|p|<k$ are ``frozen'' in the partition function, 
and mainly modes with $|p|>k$ are integrated out.
In Polchinski's approach, the IR action is kept fixed and one studies the evolution of the bare action when the arbitrary scale $k$ 
is changed. The average effective action, on the other hand, does depend on the scale $k$ and the bare effective action is fixed. 
The average effective action recovers the usual 1PI generating functional when $k\to0$, where all quantum fluctuations have the same 
weight in the partition function. 
This procedure is implemented by adding the following quadratic term to the bare action
\be\label{additional}
S_k[\phi]=\frac{1}{2}\int_p\phi_p R_k(p^2) \phi_{-p}~,
\ee 
where the smooth cut off function $R_k$ satisfies:\\
{\it(i)} $R_k\to0$ when $k\to0$, in order to recover the usual 1PI action in the deep IR;\\
{\it(ii)} $R_k(p^2)$ goes quickly to 0 for $p^2>k^2$, in order to leave undisturbed the integration over UV modes;\\
{\it(iii)} $R_k(p^2)\simeq k^2$ for $p^2\leq k^2$, which ``freezes'' IR degrees of freedom, 
by giving them the effective mass $k$.\\

Given the additional term (\ref{additional}) in the bare action, one builds the effective average action $\Gamma_k[\phi_c]$
from the average partition function
\be
Z_k[j]\equiv\int{\cal D}[\phi]\exp\left(-S[\phi]-S_k[\phi]-\int_pj_p\phi_{-p}\right)\equiv\exp(-W_k[j])~,
\ee
without forgetting to take into account the additional term (\ref{additional}) in the definition of the 
Legendre transform:
\be
\Gamma_k[\phi_c]=W_k[j]-S_k[\phi_c]-\int_xj\phi_c~.
\ee
The effect of the Legendre transform is to change the description from the functional $W$ of the source $j$ to the 
functional $\Gamma$ of the classical field $\phi_c$, assuming that there is a one-to-one mapping between $j$ and $\phi_c$.\\

This construction is similar to the introduction of the Gibbs free energy $G$, which is a Legendre transform of the energy $U$:
\be
G\equiv U-S\left(\frac{\partial U}{\partial S}\right)_V-V\left(\frac{\partial U}{\partial V}\right)_S=U-TS+PV~~,~~~~\mbox{with}~~ dU=TdS-PdV~.
\ee
The Legendre transform implies $dG=-SdT+VdP$, such that a description in terms of the variables $(S,V)$ is 
turned into a description in terms of $(T,P)$, and there is a one-to-one mapping between $(S,V)$ and $(T,P)$.\\

We note that the partition function $Z_k$ contains all the graphs of the theory, whereas $W_k$ contains the connected graphs only. The cancellation of the non-connected graphs occurs 
when taking the logarithm of $Z_k$, which can be seen with a perturbative expansion. The Legendre transform $\Gamma_k$ then contains the one-particle-irreducible graphs only, which
cannot be reduced to two graphs by cutting one internal line. As shown in Appendix A, $\Gamma_k$ corresponds to the bare action plus quantum corrections 
$\Gamma_k[\phi_c]=S[\phi_c]+{\cal O}(\hbar)$.

\vspace{0.5cm}

One can then obtain an exact functional differential equation for $\Gamma_k$, as shown here.
Keeping in mind that, after the Legendre transform, the two independent variables are $k$ and $\phi_c$, we have 
\bea
\partial_k\Gamma_k[\phi_c]&=&\frac{dW_k[j]}{dk}-\partial_kS_k[\phi_c]-\partial_k\int_xj\phi_c\nn
&=&\partial_kW_k[j]+\int_x\frac{\delta W}{\delta j}\partial_kj-\frac{1}{2}\int_p\phi_c(p) \partial_kR_k(p^2) 
\phi_c(-p)-\int_x\partial_kj\phi_c\nn
&=&\partial_kW_k[j]-\frac{1}{2}\int_p\phi_c(p) \partial_kR_k(p^2) \phi_c(-p)~.
\eea
We therefore need an expression for $\partial_kW$, which can also be written
\be
\partial_k W_k[j]=\frac{1}{2}\int_p\partial_kR_k(p^2)\left<\phi_p\phi_{-p}\right>~,
\ee
where
\be
\left<\cdots\right>\equiv\frac{1}{Z_k}\int{\cal D}[\phi](\cdots)\exp\left(-S[\phi]-S_k[\phi]-\int_pj_p\phi_{-p}\right)~.
\ee
The next step is to express the different functional derivatives of $W_k$ and $\Gamma_k$:
\bea
\frac{\delta W_k}{\delta j_{-p}}&=&\phi_c(p)\\
\frac{\delta^2W_k}{\delta j_{-p}\delta j_{-q}}&=&\phi_c(q)\phi_c(p)-\left<\phi_q\phi_p\right>\\
\frac{\delta\Gamma_k}{\delta\phi_c(p)}&=&-R_k(p^2)\phi_c(-p)-j_{-p}\\
\label{d2Gd2W}\frac{\delta^2\Gamma_k}{\delta\phi_c(p)\delta\phi_c(q)}&=&-R_k(p^2)~\tilde\delta(p+q)
-\left(\frac{\delta^2W_k}{\delta j_{-p}\delta j_{-q}}\right)^{-1}~.
\eea
Taking into account the different relations, we finally obtain the exact flow equation for $\Gamma_k$ 
\be\label{wettequa}
\partial_k\Gamma_k=\frac{1}{2}\mbox{Tr}\left\{\partial_kR_k\left(\frac{\delta^2\Gamma_k}{\delta\phi_c(p)\delta\phi_c(q)}
+R_k(p^2)~\tilde\delta(p+q)\right)^{-1}\right\}~.
\ee 
As the Polchinski equation (\ref{polchequa}), the Wetterich equation (\ref{wettequa}) is an exact self-consistent equation, but 
describes the evolution of the average effective action, instead of the bare action. Solving eq.(\ref{wettequa}) requires to
project $\Gamma_k$ onto a subspace of functionals, usually defined by the derivative expansion (\ref{gradexp}), whose convergence 
has been originally studied by Morris \cite{Morris}. We also note that, because of the cut off function $R_k$, the trace
appearing in the renormalisation equation (\ref{wettequa}) is finite and doesn't require regularisation. \\

Finally, although the IR limit $k\to0$ is, by construction, independent of the choice of cut off function $R_k(p^2)$, 
the flows do depend on $R_k(p^2)$. A particularly convenient choice is the cut off function introduced by Litim \cite{Litim}
\be\label{optcutoff}
R_k(p^2)=(k^2-p^2)\Theta(k^2-p^2)~,
\ee
which not only simplifies calculations, but also optimizes the convergence of flows, with respect to different truncations of the 
average effective action. 
Although this cut off contains a Heaviside function, it is differentiable once, and therefore is suitable for the definition of 
renormalisation flows.

\vspace{1cm}

\nin{\bf Exercise 5}: {\it Evolution of the average effective potential}\\
Neglecting the evolution of the derivative terms of the average effective action 
$$
\Gamma_k[\phi_c]=\int d^4x\left(\frac{1}{2}\partial_\mu\phi_c\partial^\mu\phi_c+U_k(\phi_c)\right)~,
$$
derive the flow equation satisfied by the potential $U_k(\phi_c)$, for the cut off function (\ref{optcutoff}).

\subsection{Example: asymptotic safety in quantum gravity}

The idea of average effective action for gravity was introduced by Reuter \cite{Reuter}. Technical details are also given 
in the original articles \cite{LauscherReuter,ReuterSaueressig1} and reviews can be found in \cite{reviews}.
These studies are motivated by the existence of a non-trivial UV fixed point for Einstein Gravity in $2+\epsilon$ dimensions \cite{Weinberg},
together with the conjecture of a non-trivial UV fixed point in higher-order-derivative gravity in 4 dimensions (see \cite{asympsafe} for
a general review of asymptotic safety in gravity).

General Relativity is perturbatively non-renormalisable, which can be understood intuitively by introducing a cut off for graviton momentum. 
If a physical quantity $P$ is calculated perturbatively, it can be expressed in the form
\be
P=P_0+GP_1+G^2P_2+G^3P_3\cdots~,
\ee
where $P_0$ is the bare quantity, $P_n$ consists in $n$-loop graphs and dots represent higher order terms in the gravitational constant $G$. Since the latter has mass dimension -2,
each term $P_n$ must be of the form $\Lambda^{2n}p_n$, where $p_n$ has the mass dimension of $P_0$. The expansion in powers of $G$ therefore seems to diverge 
when $\Lambda$ is sent to infinity
\be
P=P_0+G\Lambda^2p_1+(G\Lambda^2)^2p_2+(G\Lambda^2)^3p_3+\cdots
\ee
But one could imagine that the latter expansion could actually be re-summed to give a final result in the limit $\Lambda\to\infty$, which would 
secure the predictive power of quantum Einstein gravity. 
The studies summarized here involve Wilsonian renormalisation flows for higher-order derivative gravity, assuming that the running
average effective action lies in the functional subspace of $f(R)$ gravities.
In this context, a non-trivial UV fixed point is found, indeed suggesting asymptotic safety for gravity.\\

\nin {\bf General framework}\\

One considers a background field approach, where the total metric $g_{\mu\nu}+h_{\mu\nu}$ is decomposed into the (fixed) background metric 
$g_{\mu\nu}$ and fluctuations $h_{\mu\nu}$ which are integrated out. 
The partition function, depending on the background metric $g_{\mu\nu}$, is 
\bea
Z_k[t,\sigma,\ol\sigma]&=&\exp(-W_k[t,\sigma,\ol\sigma])\\
&=&\int{\cal D}[h,C,\ol C]\exp\Big\{-S[g+h]-S_{gf}[h]-S_{gh}[h,C,\ol C]\nn
&&~~~~~~~~~~~~~~~~~~~~~~~~-S_k[h,C,\ol C]-S_{source}[h,C,\ol C]\Big\}~,\nonumber
\eea
where
$S[g+h]$ is the bare action,
$S_{gf}[h]$ is the gauge fixing term,
$S_{gh}[h,C,\ol C]$ is the ghosts action, 
$S_k[h,C,\ol C]$ is the cut off action, and the source term is
\be
S_{source}[h,C,\ol C]=\int d^4x\sqrt{g}(h_{\mu\nu}t^{\mu\nu}+\ol\sigma_\mu C^\mu+\ol C_\mu\sigma^\mu)~.
\ee
The usual Fadeev-Popov gauge fixing term, defined in terms of the background metric, is of the form
\be
S_{gf}=\frac{1}{\alpha}\int d^4x\sqrt{g}g_{\mu\nu}F^\mu F^\nu~.
\ee
For example, the harmonic gauge is obtained with
\be
F_\mu=\kappa(\nabla^\nu h_{\mu\nu}-\frac{1}{2}\nabla_\mu h^\nu_{~\nu})~,
\ee
where $\kappa$ is a parameter with mass dimension (the covariant derivatives are taken with respect to the background metric $g_{\mu\nu}$),
and the corresponding ghost action $S_{gh}$ is quadratic in the ghosts. 
The cut off action must be quadratic in the fluctuations $h_{\mu\nu}$ and the ghosts, in order to obtain a closed self-consistent evolution 
equation for the effective action, based on the property (\ref{d2Gd2W}). The cut off action is then of the form 
\be
S_k[h,C,\ol C]=\int d^4x\sqrt{g}~h_{\mu\nu} R_k^{\mu\nu\rho\sigma} h_{\rho\sigma}+\int d^4x\sqrt{g}~\ol C_\mu Q_kC^\mu~,
\ee
where the cut off functions $R_k$ and $Q_k$ depend on the background metric $g_{\mu\nu}$ only.
The cut off function classifies modes 
to be eliminated according to the eigenvalues of the background-covariant derivatives, such that the elimination of degrees of 
freedom can be done in a ``covariant way''. This procedure is less intuitive than in flat space time though, where these eigenvalues are simply the 4-momentum. \\

The corresponding classical fields are 
\be
h_{\mu\nu}^c=\frac{1}{\sqrt{g}}\frac{\delta W_k}{\delta t^{\mu\nu}}~~,~~~~
C^\mu_c=\frac{1}{\sqrt{g}}\frac{\delta W_k}{\delta\ol\sigma_\mu}~~,~~~~
\ol C_\mu^c=\frac{1}{\sqrt{g}}\frac{\delta W_k}{\delta\sigma^\mu}~,
\ee
and the average effective action, defined on the background metric $g_{\mu\nu}$, is 
\be
\Gamma_k[h^c,C_c,\ol C_c]=W_k[t,\sigma,\ol\sigma]-S_{source}[h^c,C_c,\ol C_c]-S_k[h^c,C_c,\ol C_c]~.
\ee
We are eventually interested in the effective action as a functional of the background metric $g_{\mu\nu}$ only, 
such that we can set the classical fields to 0: $h_c=C_c^\mu=\ol C_\mu^c=0$, and look for the evolution of the relevant average effective action
\be
\Gamma^{gr}_k\equiv\Gamma_k[0,0,0]~
\ee
which depends on the background metric $g_{\mu\nu}$ in a gauge invariant way.\\

The graviton $h_{\mu\nu}$ can be decomposed into different fields
\be
h_{\mu\nu}=\ol{h}_{\mu\nu}^\bot+\frac{h}{4}g_{\mu\nu}+\nabla_\mu\xi_\nu^\bot+\nabla_\nu\xi_\mu^\bot+\left(\nabla_\mu^\rho\nabla_{\rho\nu}-\frac{1}{4}g_{\mu\nu}\nabla^2\right)\chi~,
\ee
where the spin-0 components are $h=$tr$\{h_{\mu\nu}\}$ and $\chi$;
the spin-1 component $\xi_\mu^\bot$ is transverse;
the spin-2 component $\ol h_{\mu\nu}^\bot$ is traceless and transverse. These components are orthogonal in the sense that for any $b\ne a$, the averages vanish $\left<\Phi_a\Phi_b\right>=0$,
where $\Phi_a$ is a generic notation for the different components.
Finally, if one uses an appropriate choice of gauge, $\delta^2\Gamma_k$ is diagonal in $\Phi_a$. 
If one chooses identical cut off functions for all the components $\Phi_a$ and the ghosts
(up to the tensorial structure), the evolution equation for the average effective action is of the form
\be
\partial_k\Gamma_k=\frac{1}{2}\sum_a A_a\mbox{Tr}
\left\{\left(\frac{\delta^2\Gamma_k}{\delta\Phi\delta\Phi}+R_k\right)^{-1}\partial_kR_k\right\}~,
\ee
where $A_a=1$ for bosonic fields and $A_a=-2$ for the ghosts. 
The trace is calculated using the heat kernel representation of the trace, 
for which a review can be found in \cite{heatkernel} (see also exercise 6 for the main idea).

\vspace{1cm}

\nin {\bf Exercise 6}: {\it Heat kernel representation of the trace}\\
Let $D_{x,y}$ be an inverse propagator (containing second order space time derivatives), and $\tau$ a parameter. 
By definition, the heat kernel $K(\tau,x,y)$ satisfies the diffusion equation $(\partial_\tau+D)K=0$, 
with initial condition $K(0,x,y)=\delta(x-y)$, and can formally be written $K=\exp(-\tau D)$.\\
{\it a)} Assume that $\lim_{\tau\to\infty}K=0$ and show that $D^{-1}=\int_0^\infty d\tau K$;\\
{\it b)} Show that, up to an infinite constant, any positive parameter $\lambda$ satisfies
$$
\ln\lambda=-\int_0^\infty\frac{d\tau}{\tau}\exp(-\tau\lambda)~;
$$
{\it c)} Assume that $D$ has positive eigen values, and show that
$$
\ln\mbox{det}D=-\int_0^\infty\frac{d\tau}{\tau}\mbox{tr}\{K\}~.
$$

\vspace{1cm}

\nin{\bf $f(R)$ gravity}\\

The functional space of average effective actions is infinite and it is 
impossible to take into account all the covariant operators. An approximation consists in 
projecting the running average effective action onto the functional subspace of $f(R)$ gravities \cite{f(R)}.
A proof of asymptotic safety would in principle require the complete set of different curvature terms, 
but the UV fixed point obtained for $f(R)$ gravity is a hint towards asymptotic safety.

One therefore assumes that the average effective action takes the form
\be
\Gamma^{gr}_k=\sum_{n=0}^N a_n(k)\int d^4x\sqrt{g}~R^n~,
\ee
and simplifications arise when the maximally symmetric de Sitter background is chosen, for which the Ricci scalar $R$=constant. 
This ansatz is equivalent to the local potential approximation (\ref{locpot}), and the mass dimensions of coupling constants are
$[a_n]=4-2n$. In this context, the Einstein-Hilbert action corresponds to the truncation $N=1$, defined by the relevant couplings $a_0^{EH}=2\Lambda M_{Pl}^2$ and $a_1^{EH}=M_{Pl}^2$:
\be
S_{EH}=M_{Pl}^2\int d^4x\sqrt{g}(2\Lambda+R)~,
\ee
where $M_{Pl}$ is the Planck mass and $\Lambda$ the cosmological constant. Within the Einstein truncation, one allows only the first two couplings to evolve
\be
a_0(k)\equiv 2k^4\lambda(k)/g(k)~~~,~~\mbox{and}~~~ a_1(k)\equiv k^2/g(k)~, 
\ee
where $g(k)$ is the dimensionless running gravitational constant and $\lambda(k)$ is the dimensionless running cosmological constant.
The integration of the renormalisation flows shows \cite{ReuterSaueressig1} (see Fig.2):\\
{\it(i)} a Gaussian IR fixed point $g^\star=\lambda^\star=0$;\\ 
{\it(ii)} a non-trivial UV fixed point for which $g^\star,$ and $\lambda^\star$ are of order 1.\\
Finally, this UV fixed point is shown to be stable against the choice of gauge and cut off function. Also, the renormalisation flows converge with the order $N$ of truncations in powers of 
the Ricci scalar $R$ \cite{FallsLitim}, which tends to confirm the existence of a non-trivial UV fixed point.

\begin{figure}
\epsfxsize=14cm
\centerline{\epsffile{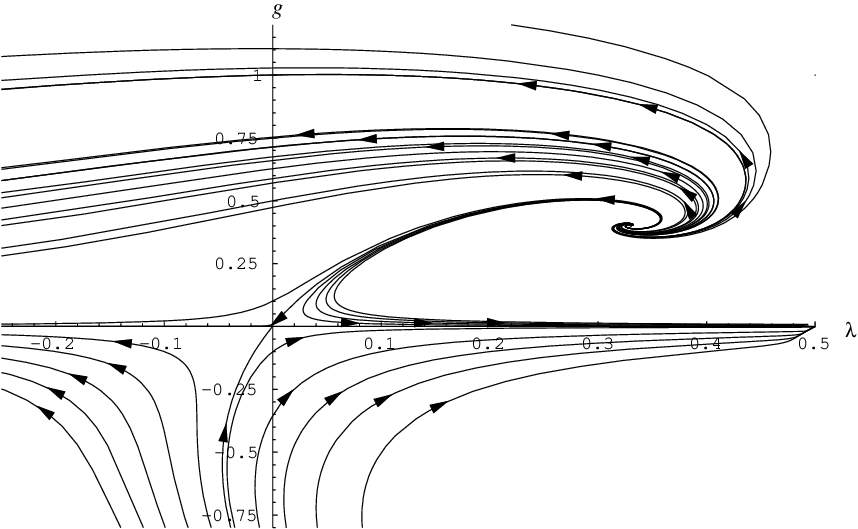}}
\caption{Renormalisation flows in the parameter space $(g,\lambda)$ for the Einstein truncation, where the arrows show the IR direction, and
the different flows correspond to different initial conditions.
Only one flow, starting with a specific initial condition in the UV, leads to the IR Gaussian fixed point, which is therefore repulsive. On the other hand, 
the non-trivial UV fixed point is attractive and any IR initial condition with $g>0$ leads to asymptotic safety.
(Figure taken from \cite{ReuterSaueressig1})}
\end{figure}

\section{Conclusion}

Three different implementations of the concept of ``blocking'' in QFT have been presented in these lectures:{\it(i)} the Wegner-Houghton, {\it(ii)} the Polchinski 
and {\it(iii)} the Wetterich approaches.
The approaches {\it(i)} and {\it(ii)} both deal with the Wilsonian effective action, although they don't lead to the same renormalisation flows. The approach {\it(iii)} deals
with the Legendre transformed effective action, and thus leads to yet another flow.
Therefore these approaches are quite different in 
their technical details, and one can question which is the most relevant in a given physical situation.

The essential point is that these approaches describe qualitatively the same Physics at large scale, and they predict the same universality classes. 
The differences arising from their technical implementation is similar to choosing different
blocks of spins, in Statistical Mechanics. Also, it has been shown in \cite{equivalent} that
there is an exact mapping between the Polchinski and the Wetterich equations, in the local potential approximation, and using the cut off function (\ref{optcutoff}).

\addcontentsline{toc}{section}{Appendix}
\section*{Appendix}

(See the introduction of \cite{AT} for the original articles on the following topics)

\addcontentsline{toc}{subsection}{A. Path integral quantisation}
\subsection*{A. Path integral quantisation}

We review here the steps of path integral quantization. Starting from the bare action $S[\phi]$, the partition function is
\be
Z[j]=\int{\cal D}[\phi]\exp\left(\frac{i}{\hbar}S[\phi]+\frac{i}{\hbar}\int_x j\phi\right)~,
\ee
where $j(x)$ is the source. The connected graph generating functional is then 
\be
W[j]=-i\hbar\ln(Z[j])~,
\ee
from which the classical field is defined as
\be\label{classicphi}
\phi_c(x)=\frac{\delta W}{\delta j(x)}.
\ee
The one-particle irreducible (1PI) graph generating functional $\Gamma[\phi_c]$ is defined as the Legendre transform of $W[j]$
\be
\Gamma[\phi_c]=W[j]-\int_x j\phi_c~,
\ee
where $j$ should be seen a functional of $\phi_c$, after inverting the definition (\ref{classicphi}).
This 1PI effective action contains all the quantum corrections of the theory, which can be seen with a saddle point approximation to 
evaluate $Z$:
\be
Z[j]=\exp\left(\frac{i}{\hbar}S[\phi_{saddle}]+\frac{i}{\hbar}\int_x j\phi_{saddle}\right)+{\cal O}(\hbar)~,\nonumber
\ee
where, by definition of the saddle point configuration $\phi_{saddle}$, 
\be
\left.\frac{\delta S}{\delta\phi}\right|_{saddle}+j=0~.
\ee
The connected graph generating functional is then
\be
W[j]=-i\hbar\ln(Z[j])=S[\phi_{saddle}]+\int_x j\phi_{saddle}+{\cal O}(\hbar)~,
\ee
and the classical field is 
\bea
\phi_c&=&\frac{\delta W}{\delta j}=\int_x\left(\left.\frac{\delta S}{\delta\phi}\right|_{saddle}+j\right)
\frac{\delta\phi_{saddle}}{\delta j}+\phi_{saddle}+{\cal O}(\hbar)\nn
&=&\phi_{saddle}+{\cal O}(\hbar)~.
\eea
From the definition of $\Gamma[\phi_c]$, one eventually obtains
\be
\Gamma[\phi_c]=S[\phi_c]+{\cal O}(\hbar)~,
\ee
such that the 1PI action is the bare action plus quantum corrections.

\addcontentsline{toc}{subsection}{B. Convexity of the 1PI effective action}
\subsection*{B. Convexity of the 1PI effective action}

From the definition of the Legendre transform $\Gamma$, one can write the equation of motion for the dressed system as
\be
\frac{\delta\Gamma}{\delta\phi_c}=\int_x\frac{\delta W}{\delta j}\frac{\delta j}{\delta \phi_c}
-\int_y\frac{\delta j}{\delta \phi_c}\phi_c-j=-j~,\nonumber
\ee
and a further derivative gives
\be\label{d2Gd2Wbis}
\frac{\delta^2\Gamma}{\delta\phi_c\delta\phi_c}=-\frac{\delta j}{\delta\phi_c}=-\left(\frac{\delta^2 W}{\delta j\delta j}\right)^{-1}~.
\ee
But one also has
\be
\frac{\delta^2W}{\delta j\delta j}=\left<\phi\right>\left<\phi\right>-\left<\phi\phi\right>~,
\ee
where
\be
\left<\cdots\right>\equiv\frac{1}{Z}\int{\cal D}[\phi](\cdots)\exp\left(-S[\phi]-\int j\phi\right)~,
\ee
which shows that the second functional derivative of $W$ is necessarily negative, as the opposite of a variance.
As a consequence, and given the relation (\ref{d2Gd2Wbis}), the second functional derivative of $\Gamma$ is positive: 
the 1PI effective action is a convex functional.
Its derivative-independent part, the 1PI effective potential, is thus a convex function.

\addcontentsline{toc}{subsection}{C. Equivalence between the Wilsonian and the 1PI effective potentials}
\subsection*{C. Equivalence between the Wilsonian and the 1PI effective potentials}

\nin The Wilsonian effective potential is defined as 
\be
\exp\left( iVU_{Wils}(\phi_0)\right)=\int{\cal D}[\phi]\delta\left(\int_x(\phi-\phi_0)\right)\exp\left( iS[\phi]\right)~,
\ee
where $V$ is the space time volume.
The Dirac distribution is then written as the Fourier transform of an exponential
\bea
\exp\left( iVU_{Wils}(\phi_0)\right) &=&\int dj\int{\cal D}[\phi]\exp\left( iS[\phi]+ij\int_x (\phi-\phi_0)\right)\nn
&=&\int dj \exp\left(iW[j]-ijV\phi_0\right)~, 
\eea
and the integration over $j$ is evaluated with the saddle point approximation, which is exact in the limit $V\to\infty$:
\be
\exp\left( iVU_{Wils}(\phi_0)\right)=\exp\left(iW[j_0]-ij_0V\phi_0\right)~,
\ee
where $j_0$ satisfies $\delta W/\delta j_0=\phi_0$, such that $\phi_0$ is the classical field corresponding to $j_0$. Finally
\be
\exp\left( iVU_{Wils}(\phi_0)\right)=\exp\left( i\Gamma[\phi_0]\right) 
=\exp\left( iVU_{1PI}(\phi_0)\right)~,
\ee
which shows the equivalence between $U_{Wils}$ and $U_{1PI}$. 
We note that this argument is valid with a Minkowski metric only, since 
it is based on the Fourier transform of the Dirac distribution.

\eject


\eject

\addcontentsline{toc}{section}{Solutions to exercises}
\section*{Solutions to exercises}

{\bf1. Critical exponents}\\
A Taylor expansion to first order around the critical point gives
$$
k\partial_k\tilde g_n\simeq f_n(g^\star)+\sum_m\left.\frac{\partial f_n}{\partial \tilde g_m}\right|_{\tilde g^\star}(\tilde g_m-\tilde g^\star_m)~,
$$
which, by definition of the fixed point, leads to
$$
k\partial_k\eta_n= \sum_m M_{nm}\eta_m~~,~~\mbox{where}~\eta_m\equiv\tilde g_m-\tilde g^\star_m~.
$$
The renormalisation equations involve dimensionless couplings, and therefore do not involve powers of $k$, such that $M$ doesn't depend on $k$. We then diagonalize $M\equiv P\Delta P^{-1}$,
where $\Delta$ is diagonal with the eigenvalues $\alpha_n$. The evolution equations in terms of $\xi\equiv P^{-1}\eta$ are then 
$$
k\partial_k\xi_n=\alpha_n\xi_n~~\mbox{(no summation over $n$)}~,
$$
with solutions
$$
\xi_n=\xi_n^0\left(\frac{k}{k_0}\right)^{\alpha_n}~.
$$

\vspace{1cm}

\nin{\bf2. Wilson-Fisher fixed point}\\
The Wegner-Houghton equation in dimension $d=4-\epsilon$ is
$$
k\partial_kU(\phi_0)=-\alpha_d k^d\ln\left(\frac{k^2+U''(\phi_0)}{k^2+U''(0)}\right)~,
$$
where 
$$
\alpha_d\equiv\frac{\hbar\Omega_d}{2(2\pi)^d}~,
$$
and $\Omega_d$ is the solid angle in dimension $d$.
Truncating the equation to $\phi^4$ and expanding to the quadratic order in the running couplings leads to
\bea
k\partial_kg_2&=&-\alpha_dk^{2-\epsilon}g_4(1-g_2/k^2)\nn
k\partial_k g_4&=&3\alpha_dk^{-\epsilon}g_4^2~.\nonumber
\eea
In dimension $d=4-\epsilon$, the field has dimension $[\phi]=1-\epsilon/2$ and the couplings $[g_{2n}]=4-2n+(n-1)\epsilon$. The 
evolution equations for the dimensionless couplings are therefore
\bea
k\partial_k\tilde g_2&=&-2\tilde g_2-\alpha_d\tilde g_4(1-\tilde g_2)\nn
k\partial_k\tilde g_4&=&-\epsilon \tilde g_4+3\alpha_d\tilde g_4^2~,\nonumber
\eea
and a non-trivial fixed point is
$$
\tilde g_4^\star=\frac{\epsilon}{3\alpha_d}~~~~\mbox{and}~~~~\tilde g_2^\star=-\frac{\epsilon}{6}~,
$$
which leads to the Gaussian fixed point $\tilde g_4^\star=\tilde g_2^\star=0$ in the limit $\epsilon\to0$.

\vspace{1cm}

\nin {\bf3.} We have
$$
\frac{\delta e^{-\Sigma_k}}{\delta\phi_p}=-e^{-\Sigma_k}\left(\frac{\delta S_k}{\delta\phi_p}+j_{-p}+\phi_{-p}Q^{-1}_k\right)~,
$$
such that
\bea
\frac{\delta^2 e^{-\Sigma_k}}{\delta\phi_p\delta\phi_{-q}}&=&e^{-\Sigma_k}\left[
\frac{\delta S_k}{\delta\phi_p}\frac{\delta S_k}{\delta\phi_{-q}}
-\frac{\delta^2 S_k}{\delta\phi_p\delta\phi_{-q}}-Q^{-1}_k\delta(p-q)\right.\nn
&&~~~~~~~~\left.+\phi_pQ^{-2}\phi_{-q}+\phi_{-p}Q^{-1}_k\frac{\delta S_k}{\delta\phi_{-q}}
+\phi_qQ^{-1}_k\frac{\delta S_k}{\delta\phi_p}\right]\nn
&&~~~~~~~+~\mbox{terms proportional to $j_p$ or $j_q$}~,\nonumber
\eea
where terms proportional to $j_p$ or $j_q$ will vanish after multiplication by $\partial_k Q^{-1}_k$. We also have
$$
Q^{-1}_k\frac{\delta(\phi_p e^{-\Sigma_k})}{\delta\phi_q}=e^{-\Sigma_k}Q^{-1}_k\left[\delta(p-q)
-\phi_p\left(\frac{\delta S_k}{\delta\phi_q}+j_{-q}+\phi_{-q}Q^{-1}_k\right)\right]~.
$$
The requested relation is finally obtained after setting $p=q$ and multiplication by $\partial_k Q^{-1}_k$ (taking into account $j_{-q}\partial_k Q^{-1}_k=0$).

\vspace{1cm}

\nin {\bf4. Quartic ansatz for the Polchinski equation}\\
From the ansatz given in the question, we have
\bea
\frac{\delta S_k}{\delta\phi_p}&=&F_k(p^2)\phi_{-p}+\frac{1}{6}\int_{qr}G_k(p,q,r)\phi_q\phi_r\phi_{-p-q-r}\nn
\frac{\delta^2 S_k}{\delta\phi_p\delta\phi_q}&=&F_k(p^2)\delta(p+q)+\frac{1}{2}\int_rG(p,q,r)\phi_r\phi_{-p-q-r}~,\nonumber
\eea
such that
$$
\frac{\delta^2 S_k}{\delta\phi_p\delta\phi_{-p}}=VF_k(p^2)+\frac{1}{2}\int_rG(p,q,r)\phi_r\phi_{-r}~,
$$
where $V=\delta(0)$ is the space time volume. One then plugs these expressions in Polchinski equation (\ref{polchequa}) and identifies the different powers of $\phi$. 
We have, on the right-hand side:\\
{\it 0th order} disregarded;\\
{\it 2nd order}
\bea
&&\frac{1}{2}\int_p\partial_kQ_k(p^2)\left[F^2_k(p^2)\phi_p\phi_{-p}-\frac{1}{2}\int_qG_k(p,-p,q)\phi_q\phi_{-q}\right]\nn
&=&\frac{1}{2}\int_p\phi_p\phi_{-p}\left[\partial_kQ_k(p^2)F^2_k(p^2)-\frac{1}{2}\int_qG_k(q,-q,p)\partial_kQ_k(q^2)\right]~;\nonumber
\eea
{\it 4th order}
$$
\frac{1}{6}\int_{pqr}\partial_kQ_k(p^2)F_k(p^2)G_k(p,q,r)\phi_p\phi_q\phi_r\phi_{-p-q-r}~;
$$
{\it 6th order} disregarded.\\
One finally finds
\bea
\partial_kF_k(p^2)&=&\partial_kQ_k(p^2)F_k^2(p^2)-\frac{1}{2}\int_qG_k(q,-q,p)\partial_kQ_k(q^2)\nn
\partial_kG_k(p,q,r)&=&4\partial_kQ_k(p^2)F_k(p^2)G_k(p,q,r)~.\nonumber
\eea

\vspace{1cm}

\nin{\bf5. Evolution of the average effective potential}\\
The second functional derivative of the average effective action is, for a constant configuration $\phi_0$,
$$
\frac{\delta^2\Gamma_k}{\delta\phi_c(p)\delta\phi_c(q)}=\Big(p^2+U_k''(\phi_c)\Big)\tilde\delta(p+q)~,
$$
such that the evolution equation gives ($V=\tilde\delta(0)$ is the space time volume)
\bea
V\partial_kU_k(\phi_c)&=&\frac{1}{2}\int\frac{d^4p}{(2\pi)^4}\frac{d^4q}{(2\pi)^4}\tilde\delta(p+q)
\partial_kR_k(p^2)\Big(p^2+R_k(p^2)+U_k''(\phi_c)\Big)^{-1}\tilde\delta(p+q)\nn
&=&\frac{\tilde\delta(0)}{2}\int\frac{d^4p}{(2\pi)^4}\frac{\partial_kR_k(p^2)}{p^2+R_k(p^2)+U_k''(\phi_c)}\nn
&=&\frac{\tilde\delta(0)}{32\pi^2}\int_0^\infty\frac{xdx~\partial_kR_k(x)}{x+R_k(x)+U_k''(\phi_c)}\nonumber
\eea
where a prime denotes a derivative with respect to $\phi_c$. for the specific choice (\ref{optcutoff})
of cut off function, we have $\partial_kR_k(p^2)=2k\Theta(k^2-p^2)$, such that 
$$
\partial_kU_k(\phi_c)=\frac{1}{32\pi^2}\frac{k^5}{k^2+U_k''(\phi_c)}~.
$$

\vspace{1cm}

\nin{\bf6. Heat kernel representation}\\
{\it a)} The propagator $D$ is independent of $\tau$, such that the integration over $\tau$ of the diffusion equation, from 0 to $\infty$, leads to
$$
[K]_0^\infty=-\delta(x-y)=-D\int_0^\infty d\tau K~,
$$
and therefore
$$
D^{-1}=\int_0^\infty d\tau K~.
$$
{\it b)} This can be obtained by taking a derivative with respect to $\lambda$. The infinite constant is the divergence at $\tau=0$, 
but which doesn't depend on $\lambda$ and so is disregarded. \\
{\it c)} Let's denote $\lambda_i$ the eigenvalues of $D$:
$$
\ln\mbox{det}D=\mbox{Tr}\ln D=\sum_i\ln\lambda_i=-\int_0^\infty\frac{d\tau}{\tau}\sum_i e^{-\tau\lambda_i}
=-\int_0^\infty\frac{d\tau}{\tau}\mbox{tr}\{K\}~.
$$

\end{document}